\documentclass[twocolumn,amsmath,amssymb,prl,showpacs]{revtex4}
\usepackage[dvips]{graphicx}
\usepackage{dcolumn}
\usepackage{bm}
\newcounter{one}
\usepackage{braket}
\usepackage[usenames]{color}
\usepackage[tight]{subfigure}

\begin{document}

\title{Generation and Eight-port Homodyne Characterization of Time-bin Qubits \\ for Continuous-variable Quantum Information Processing}

\author{Shuntaro Takeda}
\email{takeda@alice.t.u-tokyo.ac.jp}
\author{Takahiro Mizuta}
\author{Maria Fuwa}
\author{Jun-ichi Yoshikawa}
\author{Hidehiro Yonezawa}
\author{Akira Furusawa}
\email{akiraf@ap.t.u-tokyo.ac.jp}
\affiliation{Department of Applied Physics, School of Engineering, The University of Tokyo,\\ 7-3-1 Hongo, Bunkyo-ku, Tokyo 113-8656, Japan}

\date{\today}


\begin{abstract}
We experimentally generate arbitrary time-bin qubits using continuous-wave light.
The advantage unique to our qubit is its compatibility with deterministic continuous-variable quantum information processing.
This compatibility comes from its optical coherence with continuous waves, well-defined spatio-temporal mode,
and frequency spectrum within the operational bandwidth of the current continuous-variable technology.
We also demonstrate an efficient scheme to characterize time-bin qubits via eight-port homodyne measurement.
This enables the complete characterization of the qubits as two-mode states,
as well as a flexible analysis equivalent to the conventional scheme based on a Mach-Zehnder interferometer and photon-detection.
\end{abstract}

\pacs{42.50.Dv, 03.65.Wj, 03.67.Mn, 42.50.Ex}


\maketitle


There are two complementary approaches in optical quantum information processing: discrete-variables (DV) and continuous-variables (CV).
DV experiments are conducted by qubits represented by single-photon optical pulses~\cite{95Kwiat,90Rarity,99Brendel,02Marcikic,06Zavatta,04Bavichev}.
However, due to inefficient generation and imperfect detection of qubits, most of the experiments are probabilistic and require post-selection~\cite{01Tittel,12Pan}.
In contrast, CV experiments rely on the wave nature of light. They can be performed deterministically via quadrature squeezing, highly-efficient homodyne detection and feedforward operations,
at the expense of relatively low operation fidelities~\cite{98Furusawa,07Yoshikawa}.
Recently, there has emerged a ``hybrid'' approach to combine both techniques to circumvent the current limitations~\cite{11Furusawa}.
Its major advantage is deterministic operation of qubits with CV techniques;
one of the most striking examples is deterministic quantum teleportation of qubits with a CV teleporter, as is proposed in Refs.~\cite{01Ide,03Dolinska}.
The recent experiment on CV teleportation of highly non-classical optical pulses~\cite{11Lee} opens the way to this hybrid teleportation.
However, typical qubits are generated by pulse-pumped spontaneous parametric down-conversion (SPDC)~\cite{95Kwiat,90Rarity,99Brendel,02Marcikic,06Zavatta,04Bavichev}
and thus have no optical coherence with continuous-waves on which the CV teleporter is based.
Furthermore, the bandwidth of these qubits is orders of magnitude wider than the operational bandwidth of the CV teleporter (only around 10 MHz).

Here we overcome this incompatibility by generating a time-bin qubit using CW light.
This qubit consists of two temporally-separated optical pulses, described as a superposition of a photon in one pulse $\ket{1,0}$ and the other pulse $\ket{0,1}$:
$|\psi\rangle=c_0|1,0\rangle+c_1e^{i\Phi}|0,1\rangle$.
Thus far, such qubits have been prepared by pulsed lasers~\cite{99Brendel,02Marcikic,06Zavatta} and used as a key resource for various DV experiments over long distances
(e.g., quantum cryptography~\cite{00Hughes,04Marcikic}, quantum teleportation~\cite{03Marcikic}, and entanglement swapping~\cite{05Riedmatten}).
In contrast, our time-bin qubit is generated from a CW-pumped nondegenerate optical parametric oscillator (NOPO),
which is a cavity-enhanced version of the SPDC.
The NOPO cavity enhances the SPDC process only inside its resonant mode,
thereby generating qubits in a well-defined and controlled spatio-temporal mode.
The 6.2 MHz bandwidth of the resultant qubit
is within the bandwidth of the squeezing and homodyne-based feedforward operations (currently up to tens of MHz),
which mainly limits the bandwidth of the CV teleporter in Ref.~\cite{11Lee}.
Thus, our qubit is fully compatible with the teleporter as well as more advanced CV experiments based on such techniques.
Another advantage is that our qubits can be readily teleported by a single CV teleporter (such as Ref.~\cite{11Lee}) since two pulses have the same polarization,
while polarization qubits require two CV teleporters (one for each polarization)~\cite{03Dolinska}.
Such high compatibility should open the way for further hybrid protocols.

For the characterization of time-bin qubits,
we develop a scheme via eight-port homodyne measurement~\cite{84Walker,86Walker,87Walker}. 
Unlike the other schemes based on photon detection \cite{99Brendel} or regular homodyne measurement \cite{04Bavichev,06Zavatta},
our scheme provides a simple and efficient way to independently measure two pulses without varying any optical phase of the system.
This enables the reconstruction of the complete two-mode density matrices, not only in the qubit subspace spanned by $\{|1,0\rangle, |0,1\rangle\}$ but in much wider photon-number space including vacuum and multi-photon components. 
Additionally, we show that
the analysis equivalent to the conventional photon-detection scheme \cite{99Brendel} can be performed more flexibly for various detection conditions
with the same setting. 

\begin{figure}[!t]
\begin{center}
\includegraphics[width=\linewidth,clip]{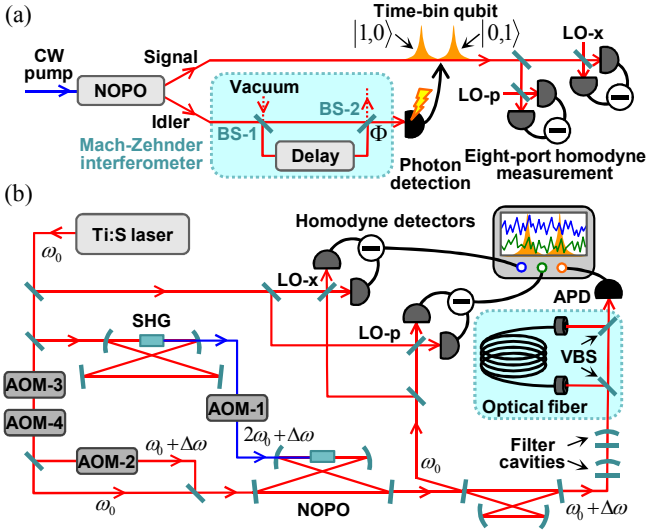}
\end{center}
\vspace{-6.6mm}
\caption{(Color online) 
(a) Schematic. (b) Experimental setup.
AOM, acousto-optic modulator;
APD, avalanche photodiode;
BS, beam splitter;
CW, continuous wave;
LO, local oscillator;
NOPO, nondegenerate optical parametric oscillator;
SHG, second harmonic generation;
and VBS, variable beam splitter.}
\label{fig:schematic}
\end{figure}


Our scheme [Fig.~\ref{fig:schematic}(a)] is based on the setup for generating single photons in Ref.~\cite{07Neergaard-Nielsen}.
In this setup, an NOPO is weakly pumped to produce correlated photon pairs in signal and idler modes written as \cite{00Loudon,11Takeda}
\begin{align}
\int dtdt^\prime C(t-t^\prime)\hat{a}^\dagger_\text{s}(t)\hat{a}^\dagger_\text{i}(t^\prime)\ket{0}_\text{s}\ket{0}_\text{i}.
\label{eq:initial_state}
\end{align}
Here, $\hat{a}^\dagger_\text{s}(t)$ and $\hat{a}^\dagger_\text{i}(t)$ are creation operators in the signal and idler mode respectively,
and $C(t-t^\prime)$ denotes the correlation function between these two modes.
A photon-detection event in the idler mode at time $t=0$ projects this state onto $\hat{a}_\text{i}^\dagger(t=0)\ket{0}_\text{i}$.
It produces a single photon state $\hat{A}^\dagger_\text{s}\ket{0}_\text{s}$,
where $\hat{A}_\text{s}^\dagger=\int dt C(t)\hat{a}^\dagger_\text{s}(t)/N$ and $N=(\int dt\left|C(t)\right|^2)^{1/2}$.
The temporal mode of this photon is defined by the mode function $f(t)=C(t)/N$,
which converges to $f_0(t)=\sqrt{\gamma}e^{-\gamma|t|}$ in the weak pumping limit ($\gamma$: NOPO bandwidth) \cite{06Molmer}.

To obtain time-bin qubits, we introduce an unbalanced Mach-Zehnder interferometer (MZI) into the idler channel as in Ref.~\cite{06Zavatta}.
This MZI manipulates the time-correlation between the signal and idler modes before photon detection,
thereby reshaping the temporal mode in which photons are generated.
This scheme originates from the theoretical proposal in Ref.~\cite{89Franson}
and the subsequent experiments in Refs.~\cite{92Brendel,93Kwiat,98Tittel},
in which an unbalanced MZI is introduced to violate Bell's inequality by generating energy-time entanglement.
We use this entanglement as a resource of time-bin qubits by detecting one of the entangled modes (the idler mode). 
This MZI reduces the generation rate by discarding one of the two output ports (it can be avoided in theory as mentioned below),
but does not degrade the purity of the generated qubits.

To model this scheme, we first introduce an auxiliary vacuum mode ``v'',
and define the beam-splitter (BS) operation on mode ``i'' and ``v'' as replacements of
$\hat{a}^\dagger_\text{i}(t)\to\tau\hat{a}^\dagger_\text{i}(t)-\rho e^{-i\Phi}\hat{a}^\dagger_\text{v}(t)$ and
$\hat{a}^\dagger_\text{v}(t)\to \rho e^{i\Phi}\hat{a}^\dagger_\text{i}(t)+\tau\hat{a}^\dagger_\text{v}(t)$,
where $\tau$, $\rho$ and $\Phi$ are the amplitude transmissivity, reflectivity and relative phase, respectively ($\tau^2+\rho^2=1$).
BS-1 $(\tau_1,\rho_1,\Phi_1=0)$ transforms Eq.~(\ref{eq:initial_state}) into
\begin{multline}
\int dtdt^\prime \big[\tau_1C(t-t^\prime)\hat{a}^\dagger_\text{s}(t)\hat{a}^\dagger_\text{i}(t^\prime)\\
-\rho_1C(t-t^\prime)\hat{a}^\dagger_\text{s}(t)\hat{a}^\dagger_\text{v}(t^\prime)\big]
\ket{0}_\text{s}\ket{0}_\text{i}\ket{0}_\text{v}.
\label{eq:state_afterBS1}
\end{multline}
By introducing a time delay $\Delta t$ in mode ``v'', the $C(t-t^\prime)$ on $\hat{a}^\dagger_\text{s}(t)\hat{a}^\dagger_\text{v}(t^\prime)$ is replaced by $C\big(t-(t^\prime-\Delta t)\big)$.
Mode ``i'' and ``v'' are then recombined by BS-2 with $(\tau_2,\rho_2,\Phi_2)$.
After this operation, the only term relevant to the photon detection on mode ``i'' has the form
\begin{multline}
\int dtdt^\prime \big[\tau_1\tau_2C(t-t^\prime)-\rho_1\rho_2 e^{i\Phi_2}C(t-t^\prime+\Delta t)\big]\\
\hat{a}^\dagger_\text{s}(t)\hat{a}^\dagger_\text{i}(t^\prime)
\ket{0}_\text{s}\ket{0}_\text{i}\ket{0}_\text{v}.
\end{multline}
By projecting this state onto $\hat{a}_\text{i}^\dagger(t=0)\ket{0}_\text{i}$ and tracing out the unused mode ``v'', we obtain
\begin{align}
\int dt\left[\tau_1\tau_2C(t)-\rho_1\rho_2e^{i\Phi_2}C(t+\Delta t)\right]\hat{a}^\dagger_\text{s}(t)\ket{0}_\text{s}.
\label{eq:time-bin_state}
\end{align}
When $\Delta t$ is sufficiently long compared to the photon coherence time ($\sim1/\gamma$) between the signal and idler modes, the two modes defined by
$\hat{A}_1^\dagger=\int dt C(t)\hat{a}^\dagger_\text{s}(t)/N$ and
$\hat{A}_2^\dagger=\int dt C(t+\Delta t)\hat{a}^\dagger_\text{s}(t)/N$ can be regarded as orthogonal.
Thus, from Eq.~(\ref{eq:time-bin_state}), the two-mode state can be described as
$\ket{\psi}_{12}\propto\tau_1\tau_2\ket{1}_1\ket{0}_2-\rho_1\rho_2e^{i\Phi_2}\ket{0}_1\ket{1}_2$.
The delay ($t^\prime\to t^\prime-\Delta t$) in the idler mode induces earlier time-bin ($t\to t+\Delta t$) in the signal mode,
and the two time-bins constitute a single time-bin qubit.
The coefficients of the qubit are determined by the splitting ratio of both BS-1 and BS-2, as well as the recombining phase at BS-2;
we experimentally show that they are arbitrarily tunable.
In the weak pumping regime, the mode function of each time-bin is written as $f_1(t)=\sqrt{\gamma}e^{-\gamma|t|}$ and $f_2(t)=\sqrt{\gamma}e^{-\gamma|t+\Delta t|}$.
Note that, when the photon detection is performed on the other output port of BS-2 instead,
$\ket{\psi^\prime}_{12}\propto\tau_1\rho_2\ket{1}_1\ket{0}_2+\rho_1\tau_2e^{i\Phi_2}\ket{0}_1\ket{1}_2$
is obtained.
By adding photon-detection events of this port and performing an appropriate unitary transformation thereafter,
the MZI-induced reduction of the generation rate is avoidable, though not experimentally shown here.


Our detailed experimental setup is shown in Fig.~\ref{fig:schematic}(b).
Part of the output beam of the Ti:Sapphire laser (wavelength: 860 nm) is
frequency doubled by a second-harmonic generation cavity (frequency: $\omega_0\to2\omega_0$).
Its frequency is then shifted with an acousto-optic modulator (AOM-1) by 590 MHz ($2\omega_0\to2\omega_0+\Delta\omega$), which corresponds to the free spectrum range (FSR) of our NOPO.
10 mW of this beam pumps the NOPO [half-width at half-maximum (HWHM): 6.2 MHz ($=\gamma/2\pi$)], 
producing signal ($\omega_0$) and idler ($\omega_0+\Delta\omega$) photons.
Two weak coherent beams ($\omega_0$, $\omega_0+\Delta\omega$) are also injected into the NOPO to lock the subsequent system.
Both beams are alternatively blocked and unblocked by AOM-3 and 4 at a rate of 2 kHz.
The signal and idler modes are spatially separated by a cavity (HWHM: 22.2 MHz, FSR: 1150 MHz).
The idler mode is then injected into a MZI after filtering out irrelevant NOPO modes via two filter cavities.
In the MZI, part of the idler mode is split off by a variable BS (VBS), composed of a half-wave plate and a polarizing BS.
The split-off part is transmitted through a 50m optical fiber (PM780-HP, Thorlabs), and is then recombined with another VBS.
Here the delay is set to $\Delta t=242$ ns so that the overlap between two time-bins is negligible ($\int f_1(t)f_2(t)dt\sim 0.08\%$).
The optical path length difference is passively stabilized by surrounding the fiber with heat insulating material,
as well as actively locked to a target value with piezo actuators.
One output port of the MZI is sent to an avalanche photodiode (APD) to herald time-bin qubits,
whereas the other is monitored by a photodetector to phase lock the MZI.


\begin{figure}[!t]
\begin{center}
\includegraphics[width=\linewidth,clip]{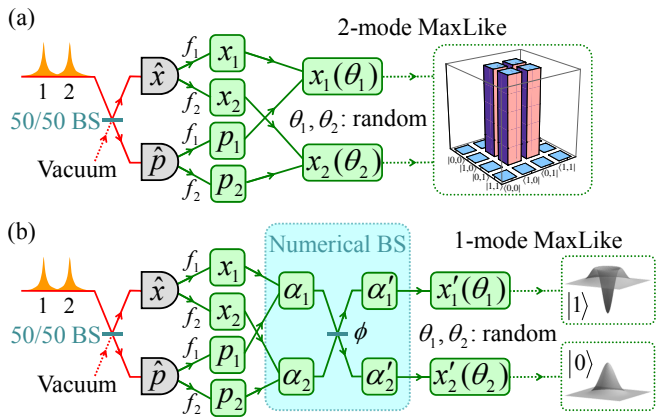}
\end{center}
\vspace{-6mm}
\caption{(Color online) Eight-port homodyne analysis of time-bin qubits.
BS, beam splitter; and MaxLike, maximum likelihood algorithm.
(a) Reconstruction of two-mode density matrices.
(b) The analysis equivalent to the conventional photon-detection scheme.}
\label{fig:analysis}
\end{figure}

In general, complete homodyne characterization of two-mode states requires two-mode quadrature measurements at various phase sets $(\theta_1,\theta_2)$ \cite{04Bavichev}.
One possible way to carry out this in our case is
to spatially separate two time-bins and then perform a homodyne-measurement on each bin with careful control of the local oscillator's phase.
However, since two bins have the same polarization and are temporally close to each other, such a separation requires high-speed switching with high precision.
In Ref.~\cite{06Zavatta}, only the relative phase of two time-bins are scanned instead of fully scanning local oscillators' phases, but this scheme is inapplicable when characterizing unknown qubit states.
Here, we propose a more efficient scheme via eight-port homodyne measurement~\cite{84Walker,86Walker,87Walker} where no optical phase scanning is required.
First, both time-bins are split with a 50/50 BS for two homodyne measurements for orthogonal quadratures $\hat{x}$ and $\hat{p}$.
By multiplying $f_1(t)$ and $f_2(t)$ to each homodyne photocurrent,
we can obtain the quadratures $(x_1, p_1, x_2, p_2)$ with extra vacuum noise from the unused port of the BS [Fig.~\ref{fig:analysis}(a)];
this vacuum noise is the inevitable consequence of the simultaneous measurement of conjugate variables $\hat{x}$ and $\hat{p}$~\cite{65Arthurs}.
From this set, quadratures at any phase $\theta$ can be calculated as $x_j(\theta)=x_j\cos\theta+p_j\sin\theta$ ($j=1,2$).
Therefore, from one set, we can randomly select a phase set $(\theta_1,\theta_2)$ and construct two independent tomography data sets
$\big(x_1(\theta_1), x_2(\theta_2)\big)$ and $\big(x_1(\theta_1+\pi/2), x_2(\theta_2+\pi/2)\big)$
which refer to orthogonal quadratures.
In the experiment, 100000 quadrature sets are acquired and
200000 tomography data sets are constructed.
The maximum likelihood algorithm~\cite{04Lvovsky} is then used to reconstruct two-mode density matrices in the Fock basis:
$\hat{\rho}=\sum_{k,l,m,n=0}^{\infty}\rho_{klmn}\ket{k,l}\bra{m,n}$.
Here we compensate only the extra vacuum noise added in the dual-homodyne measurement,
and not other experimental imperfections.
The scheme mentioned above can be easily extended to single- or higher-mode states,
which shows that eight-port homodyne measurement is a useful tool for characterizing multi-mode states.
Note that, in the present setting,
the eight-port homodyne measurement is directly measuring the two-mode $Q$ function of a qubit~\cite{03Freyberger,03Leonhardt}.
Once the $Q$ function is estimated from the collected data,
the corresponding density matrix can be calculated in theory~\cite{10Leonhardt}.
Instead, we have mapped the measured data $(x_1, p_1, x_2, p_2)$ to $\big(\theta_1, x_1(\theta_1), \theta_2, x_2(\theta_2)\big)$,
from which we can directly calculate the density matrix using the standard maximum likelihood algorithm~\cite{04Lvovsky}.

\begin{figure}[!b]
\begin{center}
\includegraphics[width=0.93\linewidth,clip]{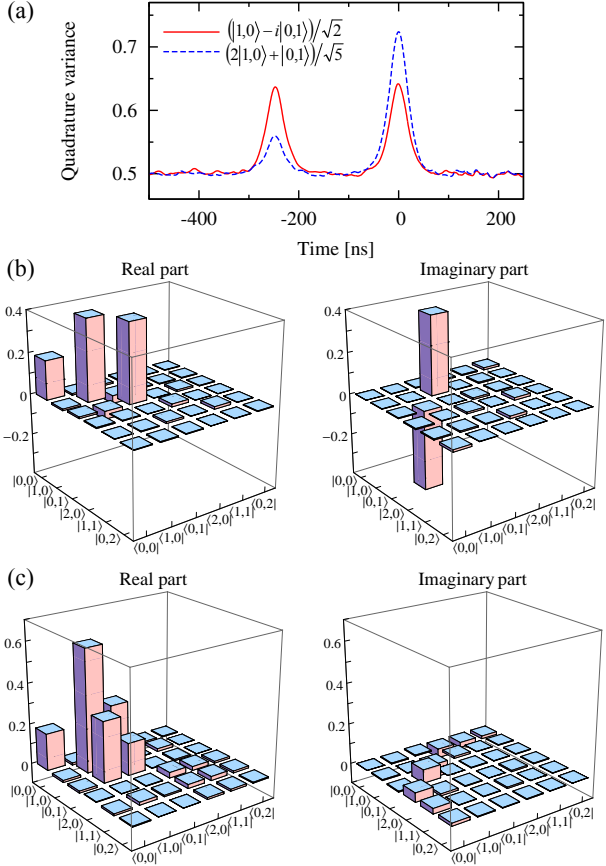}
\end{center}
\vspace{-7mm}
\caption{(Color online) Experimental results. Quadrature variance at each time point (a) and density matrices of (b): $(|1,0\rangle-i|0,1\rangle)/\sqrt2$ and
(c): $(2|1,0\rangle+|0,1\rangle)/\sqrt5$.}
\label{fig:two-mode_density}
\end{figure}

We generated eight types of qubits: $(\ket{1,0}+e^{i\Phi}\ket{0,1})/\sqrt2$ and $(2\ket{1,0}+e^{i\Phi}\ket{0,1})/\sqrt5$ with $\Phi=0,\pi,\pm\pi/2$.
Figure \ref{fig:two-mode_density}(a) shows the quadrature variance at each time point calculated from the 100000 quadrature traces where the vacuum variance is normalized to $1/2$ ($\hbar=1$).
Two time-bins appear as the two peaks of the trace, and the peak height above the vacuum noise level corresponds to the probability of existing photons~\cite{07Neergaard-Nielsen}.
Experimental density matrices in Fig.~\ref{fig:two-mode_density}(b,c)
show not only the qubit submatrix spanned by $\{\ket{1,0},\ket{0,1}\}$
but also the vacuum and multiphoton contributions.
Ideally, generated states have components only in the qubit submatrix.
In practice, the density matrices show $18\pm1\%$ of a vacuum ($\rho_{0000}$),
$77\pm1\%$ of a qubit ($\rho_{1010}+\rho_{0101}$), and $5\pm1\%$ of multiphoton components ($1-\rho_{0000}-\rho_{1010}-\rho_{0101}$).
The ratio of diagonal elements $\rho_{1010}$ to $\rho_{0101}$ is equal to the ratio of $\ket{1,0}$ to $\ket{0,1}$,
whereas the off-diagonal elements $\rho_{1001}$ and $\rho_{0110}$ demonstrate the superposition of $\ket{1,0}$ and $\ket{0,1}$ at the target phase.
The encoded quantum information can be read out by extracting and renormalizing the qubit submatrix.
The average fidelity of each submatrix with its target state is $0.989\pm0.004$,
showing the highly precise qubit preparation.

The $18\%$ vacuum contribution is well explained by the estimated loss of $1-\eta_\text{all}=16\%$,
where $\eta_\text{all}=\eta_\text{NOPO}\eta_\text{vis}^2\eta_\text{pr}\eta_\text{det}\eta_\text{APD}$ is the overall efficiency.
Here $\eta_\text{NOPO}=0.98$ is the escape efficiency of the NOPO,
$\eta_\text{vis}=0.98$ the average mode-matching visibility at eight-port homodyne measurement,
$\eta_\text{pr}=0.96$ the propagation efficiency from the NOPO to the homodyne detectors,
$\eta_\text{det}=0.95$ the average detection efficiency given by the quantum efficiency of photodiodes and the electronic noise,
and $\eta_\text{APD}=(\zeta_\text{tot}-\zeta_\text{dark})/\zeta_\text{tot}=0.98$ the purity of the photon-detection event given by
the total count rate $\zeta_\text{tot}=5800 \text{ s}^{-1}$ and the dark count rate $\zeta_\text{dark}=80 \text{ s}^{-1}$.


\begin{figure}[!t]
\begin{center}
\includegraphics[width=\linewidth,clip]{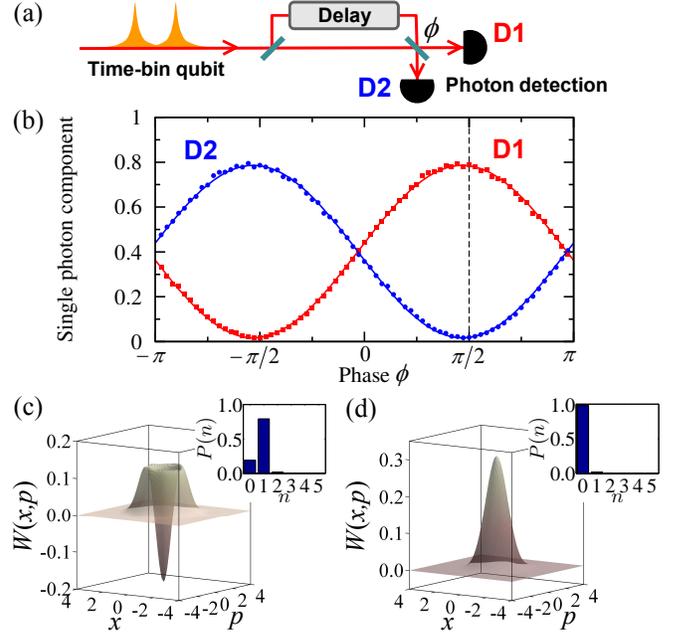}
\end{center}
\vspace{-6mm}
\caption{(Color online) (a) Conventional photon-detection scheme.
(b) The $\phi$ dependence of single photon components of two modes which are subject to photon detections D1 and D2.
(c,d): Experimental Wigner functions $W(x,p)$ and photon-number distributions $P(n)$ at $\phi=\pi/2$.
}
\label{fig:disentangle}
\end{figure}

Our eight-port homodyne setup also enables the analysis equivalent to the conventional detection scheme in Fig.~\ref{fig:disentangle}(a),
where two time-bins are recombined via a MZI and then the existence of photons at each output mode is monitored by photon detections D1 and D2.
In this case, D1 and D2 correspond to the projection onto orthogonal qubit bases, and if the BS parameters are chosen properly,
photons arrive at only one of the two.
In our system, this BS can be realized by analytically mixing measured quadratures of two time-bins after the measurement [Fig.~\ref{fig:analysis}(b)].
For a quadrature data set $(x_1, p_1, x_2, p_2)$,
the complex amplitude of each mode can be written as $\alpha_j=(x_j+ip_j)/\sqrt2$ ($j=1,2$).
When the BS recombines these two time-bins with $(\tau^\prime,\rho^\prime,\phi)$,
the output amplitude $(\alpha_1^\prime, \alpha_2^\prime)$ can be calculated as
$\alpha_1^\prime=\tau^\prime \alpha_1+\rho^\prime e^{i\phi}\alpha_2$ and
$\alpha_2^\prime=-\rho^\prime e^{-i\phi}\alpha_1+\tau^\prime \alpha_2$.
Finally the output quadratures at any given phase $\theta$ can be calculated from these amplitudes
as $x_j^\prime(\theta)=(\alpha_j^\prime e^{-i\theta}+\alpha_j^{\prime*}e^{i\theta})/\sqrt2$.
Therefore, we can compose a data set $(x_1^\prime(\theta_1),x_2^\prime(\theta_2))$ for any desired BS parameters $(\tau^\prime,\rho^\prime,\phi)$ and any phase set $(\theta_1,\theta_2)$.
As a result, the density matrix of each mode subject to D1 or D2 can be reconstructed
by maximum likelihood algorithm compensating the extra vacuum noise.
In the special case of choosing $(\tau^\prime,\rho^\prime,\phi)=(c_0,c_1,-\Phi)\text{ or }(c_1,c_0,-\Phi+\pi)$
for an initial time-bin qubit $\ket{\psi}_{12}=c_0\ket{1}_1\ket{0}_2+c_1 e^{i\Phi}\ket{0}_1\ket{1}_2$,
the two modes would theoretically be decomposed into a single photon state and a vacuum state.
In our eight-port homodyne scheme, the whole set $\{(\tau^\prime,\rho^\prime,\phi)\}$ of the D1-D2 measurement results can be obtained from a single quadrature data set.
Some related discussions on this photon-vacuum decomposition were made in terms of the cross section of two-mode Wigner function in Refs.~\cite{04Bavichev,06Zavatta}.

We analyzed the data set of the qubit $|\psi\rangle=(|1,0\rangle-i|0,1\rangle)/\sqrt2$
for a fixed $\tau^\prime=\rho^\prime=1/\sqrt2$ and varying $\phi$.
The $\phi$ dependence of the single photon component of D1 and D2 [Fig.~\ref{fig:disentangle}(b)] shows the fringe visibility of $96\pm2\%$.
It reaches its minimum and maximum around $\pm\pi/2$, demonstrating the generation of the target qubit state.
The reconstructed states at $\phi=\pi/2$
are explicitly decomposed into a single photon and a vacuum.
Figure~\ref{fig:disentangle}(c) shows $P_\text{a}(1)=0.79$ of a single photon component
and a strong negativity of Wigner function $W_\text{a}(0,0)=-0.187$.
Figure~\ref{fig:disentangle}(d) is an almost pure vacuum with $P_\text{b}(0)=0.98$.
$P_\text{b}(1)=0.02$ of a photon still remains, possibly because the BS parameters $(\tau^\prime,\rho^\prime,\phi)$ are not optimal
and the phase $\Phi$ of the qubit fluctuated during the measurement.


In summary,
we have generated arbitrary time-bin qubits in a well-defined spatio-temporal mode by using a CW-pumped NOPO.
The generated qubit is compatible with the current CV operations,
thereby providing an essential tool for hybrid approaches in quantum information processing.
Our eight-port homodyne scheme efficiently reconstructed the full two-mode density matrices of the qubits,
and also enabled the analysis equivalent to the conventional photon-detection scheme by a numerical BS operation.
This characterization scheme is universal, and applicable to
single- or multi-mode states encoded in different temporal modes~\cite{06Zavatta,02Marcikic,99Brendel}, spatial modes~\cite{90Rarity,04Bavichev} or polarization modes~\cite{95Kwiat}.

\begin{acknowledgments}
This work was partly supported by the SCOPE program of the MIC of Japan, PDIS, GIA, G-COE, APSA, and FIRST commissioned by the MEXT of Japan, and ASCR-JSPS.
\end{acknowledgments}


\end{document}